\documentclass{IEEEtran4PSCC}

\usepackage{cite}

\ifCLASSINFOpdf
   \usepackage[pdftex]{graphicx}
\else
   \usepackage[dvips]{graphicx}
\fi
\usepackage[cmex10]{amsmath}
\usepackage{algorithmic}

\usepackage{array}
\makeatletter
\let\old@ps@headings\ps@headings
\let\old@ps@IEEEtitlepagestyle\ps@IEEEtitlepagestyle
\def\psccfooter#1{%
    \def\ps@headings{%
        \old@ps@headings%
        \def\@oddfoot{\strut\hfill#1\hfill\strut}%
        \def\@evenfoot{\strut\hfill#1\hfill\strut}%
    }%
    \def\ps@IEEEtitlepagestyle{%
        \old@ps@IEEEtitlepagestyle%
        \def\@oddfoot{\strut\hfill#1\hfill\strut}%
        \def\@evenfoot{\strut\hfill#1\hfill\strut}%
    }%
    \ps@headings%
}
\makeatother

\usepackage{bbm}
\usepackage{amssymb}
\usepackage{graphicx}
\usepackage{tikz}
\usepackage{pgfplots} 
\usepackage{pgfgantt}
\usepackage{pdflscape}
\pgfplotsset{compat=newest} 
\pgfplotsset{plot coordinates/math parser=false}
\newlength\fwidth
\newlength\fheight
\usetikzlibrary{shapes,arrows}
\usepackage{mathtools}
\usepackage{adjustbox}
\usepackage{eurosym}

\usepackage[hyperindex=true, %
  pdftitle={Tuning and Testing an Online Feedback Optimization Controller to Provide Curative Distribution Grid Flexibility},%
  pdfauthor={Lukas Ortmann, Fabian Böhm, Florian Klein-Helmkamp, Andreas Ulbig, Saverio Bolognani, Florian Dörfler},%
  colorlinks=true,%
  pagebackref=false,%
  plainpages=false,%
  pdfpagelabels,%
  linkcolor=black,%
  citecolor=black,%
  filecolor=black,%
  urlcolor=black%
  ]{hyperref}

\usepackage{acronym}
\acrodef{OFO}{Online Feedback Optimization}
\acrodef{FPU}{Flexibility Providing Units}
\acrodef{PCC}{point of common coupling}
\acrodef{QP}{quadratic problem}

\renewcommand{\r}{\textcolor{black}}

\begin{document}
%
\title{Tuning and Testing an Online Feedback Optimization Controller to Provide Curative Distribution Grid Flexibility}

\author{
\IEEEauthorblockN{Lukas Ortmann, Saverio Bolognani, Florian Dörfler}
\IEEEauthorblockA{Automatic Control Laboratory\\ETH Zurich, Switzerland}
\and
\IEEEauthorblockN{Fabian Böhm, Florian Klein-Helmkamp, Andreas Ulbig}
\IEEEauthorblockA{IAEW at RWTH Aachen\\
Aachen, Germany
}
}

\maketitle

\begin{abstract}
\r{Due to more volatile generation, flexibility will become more important in transmission grids. One potential source of this flexibility can be distribution grids.} A flexibility request from the transmission grid to a distribution grid then needs to be split up onto the different \ac{FPU}s in the distribution grid. One potential way to do this is \ac{OFO}. \ac{OFO} is a new control method that steers power systems to the optimal solution of an optimization problem using minimal model information and computation power. This paper will show how to choose the optimization problem and how to tune the \ac{OFO} controller. Afterward, we test the resulting controller on a real distribution grid laboratory and show its performance, its interaction with other controllers in the grid, and how it copes with disturbances. Overall, the paper makes a clear recommendation on how to phrase the optimization problem and tune the \ac{OFO} controller. Furthermore, it experimentally verifies that an \ac{OFO} controller is a powerful tool to disaggregate flexibility requests onto \ac{FPU}s while satisfying operational constraints inside the flexibility providing distribution grid.
\end{abstract}


\section{Introduction}

\begin{figure}
    \centering
    \setlength\fwidth{\columnwidth}
        \setlength\fheight{8cm}
        \begin{adjustbox}{max width=\columnwidth}
	    \begin{tikzpicture}
\begin{axis}[
	x tick label style={
		/pgf/number format/1000 sep={}, font=\scriptsize},
    y tick label style={/pgf/number format/1000 sep={}},
	ylabel= {Cost [Mio \euro/a ]},
    xlabel= {Year},
	enlargelimits=0.05,
	ybar interval=0.7
]
\addplot[blue, fill=blue] 
	coordinates {(2013,214.8) (2014,436.1) 
		(2015,1141.2) (2016,893.3)
        (2017,1448.2) (2018, 1353.2)
        (2019,1178.2) (2020, 1398.6)
        (2021,2285.7) (2022,4248)
        (2023,0)};
\end{axis}
\end{tikzpicture}
        \end{adjustbox}
    \caption{Cost for redispatch in Germany. The data is taken from~\cite{redispatch_cost}.}
    \label{fig:congestion_management}
\end{figure}
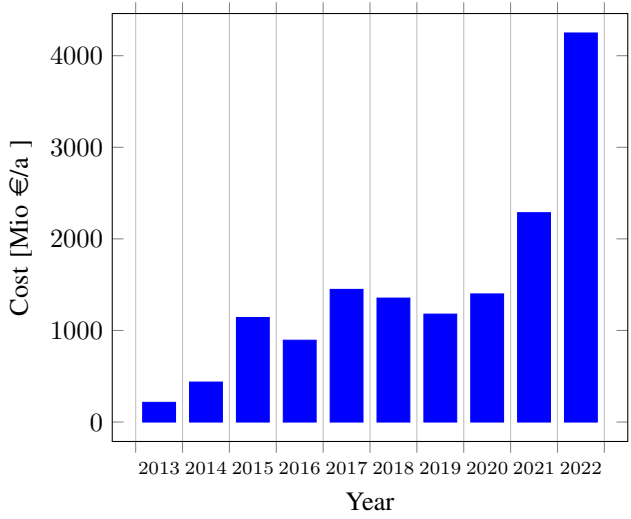

Curative measures are an important tool for grid operators to guarantee the safe operation of their grid~\cite{Innosys,kolster2020contribution}. Lately, such redispatch actions have been used increasingly often \r{as can be seen in} Figure~\ref{fig:congestion_management}. Curative actions can either be provided by a single power plant or an aggregation of many small entities, so-called \ac{FPU}s. Such an aggregation can for example be a distribution grid. To be able to provide a curative action with a distribution grid, it is important to estimate the available flexibility of the distribution grid~\cite{silva2018estimating}. It is equally important to then have tools that can disaggregate the curative action onto the potential heterogeneous group of \ac{FPU}s in the distribution grid. This coordination needs to be fast and it is also important that the operation constraints in that distribution grid are satisfied at all times. One tool to disaggregate the curative action is by defining an optimization problem whose solution is the setpoints to the \ac{FPU}s~\cite{fruh2022coordinated}. There are then several ways to solve such an optimization problem. Option~1 is to solve the optimization offline and in a feedforward approach using a model of the distribution grid. However, such optimization approaches can be computationally intense, and more importantly they heavily rely on model knowledge which means they are prone to model mismatch. Furthermore, if the entity providing the curative action is a distribution grid then the optimization also needs to be robust against changing consumption and generation within that grid. One way to achieve this robustness is to use robust optimization as was done in these papers~\cite{kolster2022providing,chen2021leveraging}. Unfortunately, such approaches are even more computationally intense and to guarantee robustness they do not utilize the available flexibility of a distribution grid to its full extent. To circumvent these difficulties, the authors of~\cite{klein2023providing} have proposed to use \ac{OFO}. This method solves the same optimization problem as model-based approaches but incorporates feedback from the grid through measurements which makes it robust to model mismatch and changing consumption and production in that grid~\cite{hauswirth2021optimization}. This allows us to achieve robustness while still using the full available flexibility of the distribution grid. Furthermore, the needed model information is significantly smaller~\cite{ortmann2020experimental}. \ac{OFO} controller have been used in a large number of papers in simulations~\cite{olives2022model,nowak2020measurement_PVSC,gan2016online,dall2016optimal,olives2023holistic,tang2020measurement,bernstein2019real,zhan2023fairness} and hardware-in-the-loop experiments~\cite{wang2020performance,padullaparti2021peak}. Experiments using \ac{OFO} on a real power grid setup were done by~\cite{ortmann2020experimental,ortmann2020fully,reyes2018experimental,kroposki2020autonomous,kroposki2020good,ortmann2023deployment}. However, none of these papers deal with the use case of curative actions/ curative distribution grid flexibility nor the tuning of \ac{OFO} controllers. Therefore, this paper will analyze in detail how \ac{OFO} can be used to disaggregate curative actions onto \ac{FPU}s. Furthermore, the tuning of an \ac{OFO} controller is discussed in detail.

Whenever optimization tools, either model-based or \ac{OFO}, are used to disaggregate curative actions onto \ac{FPU}s, the first challenge is to define a suitable optimization problem. A second challenge is to tune the algorithm that solves the optimization problem, may it be a model-based approach or \ac{OFO}. To address the first challenge, we compare two fundamentally different ways to define an optimization problem in this context and give a clear recommendation of which formulation is the superior one. To address the second challenge, we analyze the convergence behavior of an \ac{OFO} controller in this application context and showcase how it needs to be tuned such that a heterogeneous group of \ac{FPU}s accurately provides the requested curative action. Finally, we validate the developed and tuned \ac{OFO} controller on a real distribution grid. This enables us to show the robustness of \ac{OFO} against voltage fluctuations originating from the upper-level grid, other control structures in the grid that are operating independently from our controller, and changing production and consumption in the distribution grid. Also, this enables us to show that our \ac{OFO} controller can accurately and quickly disaggregate a curative action request onto a heterogeneous group of \ac{FPU}s in a distribution grid while guaranteeing the satisfaction of the operation limits in that distribution grid.

\section{Laboratory Setup}

\begin{figure}
    \centering
    \includegraphics[width=\columnwidth]{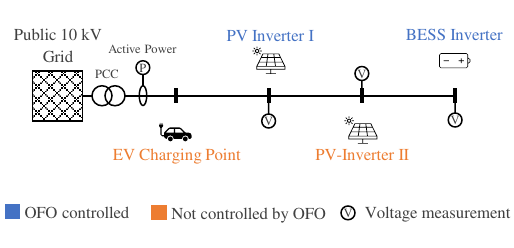}
    \caption{Laboratory setup used for the experimental validation. PV inverter I is connected to bus 3, PV inverter II is connected to bus 1, and the BESS inverter is connected to bus 2. The figure is taken from~\cite{klein2023providing}.}
    \label{fig:topology}
\end{figure}

We implement our \ac{OFO} controllers on a laboratory distribution grid located at RWTH Aachen University to show how they behave in a real grid. The grid is operated at 400V and connected to the public medium voltage grid at the \ac{PCC}. The grid consists of 850m of cables that span a long radial grid to which we connect two PV inverters, a battery energy storage system (BESS) inverter, and an EV charging point. The \ac{OFO} controller derives active and reactive power setpoints for one of the PV inverters and the BESS inverter, see Figure~\ref{fig:topology}. The EV charging point and the second PV inverter are not part of the \ac{OFO} control. To be more precise, the \ac{OFO} controller does not even know they are connected to the grid. Voltage measurements ($v$) are taken throughout the grid and the power flow is measured at the \ac{PCC} ($p_{PCC}$). The \ac{OFO} controller has access to this active power measurement and the voltages at the devices it derives the setpoints for. The sampling time of the \ac{OFO} controller is 5~seconds and the control setup is implemented using an existing software interface to the laboratory grid \cite{hacker2021framework,schmidtke2022evaluation}.
The outlined configuration inherently encompasses multiple potential external interferences that impact the OFO controller while coordinating flexibility. These interferences consist of the following:
a) fluctuating voltage levels at the \ac{PCC}, b) reactive power injections and consumptions of inverters that are not controlled by our \ac{OFO} controller, c) variable power consumption stemming from the EV charging station.
Throughout the experiments, conducted as part of this study, the robustness of the OFO algorithm was evaluated against any of the aforementioned disturbances. Elaborated details regarding the outcomes of these experiments can be found in the subsequent sections.

\section{Introduction to Online Feedback Optimization}

In this section, we give a quick introduction to \ac{OFO}.
Consider a system with inputs $u$, disturbances $d$, and outputs $y=h(u,d)$, where $h(\cdot,\cdot)$ describes the behavior of the system. In our use case, the input is $u = \begin{bmatrix} p\\q\end{bmatrix}$, where $p$ and $q$ are the active and reactive power commands to the \ac{FPU}. The output is $y=\begin{bmatrix} v\\p_{PCC}\end{bmatrix}$, where $v$ are the measured voltage magnitudes in the grid and $p_{PCC}$ is the active power flow over the \ac{PCC}.
The goal of \ac{OFO} is to bring the inputs $u$ of the system to an optimal operating point $u^\star$. This optimal point is defined through an optimization problem for example:
\begin{equation} \label{eq:general_op_problem}
	\begin{aligned}
		\min_{u,y} \quad & \Phi(u,y) \\
		\textrm{s. t.} \quad & u_{min} \leq u \leq u_{max} \\
        & y_{min} \leq y \leq y_{max} \\
        &y=h(u,d)
\end{aligned}
\end{equation}

\ac{OFO} is a feedback control method that uses optimization algorithms to solve such optimization problems.
It is important to note that, \ac{OFO} does not calculate $u^\star$ offline and based on a model but iteratively and online. This means the inputs \r{$u(k)$}, which are applied to the real system, are changed over time until they have converged to the solution of the optimization problem $u^\star$. This makes the approach robust to model mismatch, computationally light, and nearly model free~\cite{hauswirth2021optimization}.\\
The type of \ac{OFO} controller we use in this paper is based on projected gradient descent. It consists of the integral controller
${u(k+1) = u(k) + \sigma(u,d,y_m)}$
and an underlying \ac{QP}
\begin{equation} \label{eq:ofostepcalculation}
	\begin{aligned}
		\sigma(u,d,y_m) \coloneqq \arg &\min_{w \in \mathbb{R}^{p}} \, \| w + G^{-1} H(u,d)^T\nabla \Phi(u,y_m)\|^{2}_G\\
		\textrm{s.t.} \quad & u_{min} \leq u(k) + w < u_{max}\\
          \quad & y_{min} \leq y_m(k) + \nabla_u h(u,d)w < y_{max}
\end{aligned}
\end{equation} 
with $w = \begin{bmatrix}
    \Delta p\\ \Delta q
\end{bmatrix}$, the tuning matrix $G$, and $H(u,d)^T=[I_n \, \nabla_u h(u,d)^T]$ where $n$ is the number of entries in $u$ and $()^T$ denotes the transpose.
The term $\nabla_u h(u,d)$ is the gradient of $y$ with respect to $u$. It is a linear approximation of $h(\cdot,\cdot)$ and describes how a change in $u$ affects $y$. In this paper, it describes how changes in $p$ and $q$ change the voltages $v$ and the power flow $p_{PCC}$. Therefore $\nabla_u h(u,d)$ consists of power transfer distribution factors and other sensitivities. We define $H_p = \begin{bmatrix} \nabla_p p_{PCC}(p,q,d)\\\nabla_q p_{PCC}(p,q,d)\end{bmatrix}\approx\begin{bmatrix} \nabla_p p_{PCC}(p,q,d)\\0\end{bmatrix}$ and $H_v = \begin{bmatrix} \nabla_p v(p,q,d)\\\nabla_q v(p,q,d)\end{bmatrix}$. Note that, both $H_p$ and $H_v$ can be approximated by a constant vector or matrix, respectively. We calculate such approximations offline for a fixed initial operating point using a model of the grid and use these same sensitivity values throughout the paper and all experiments. This is possible because \ac{OFO} is robust against model mismatch in these sensitivities~\cite{ortmann2020experimental}.\\
\r{Overall, the setup is as depicted in the block diagram that can be seen in Figure~\ref{fig:block_diagram}.
\begin{figure}
    \centering
    \includegraphics[width=\columnwidth]{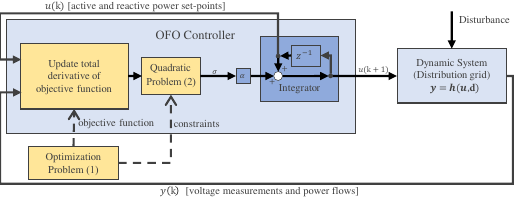}
    \caption{Block diagram of an \ac{OFO} controller. The OFO controller iteratively changes $u$ until the solution of the optimization problem is reached. The figure is adapted from~\cite{klein2023providing}.}
    \label{fig:block_diagram}
\end{figure}
The \ac{OFO} controller is implemented in a centralized location. It receives voltage measurements $v$ and a measurement of the active power flow over the \ac{PCC} $p_{PCC}$. Based on these measurements the \ac{OFO} controller updates the active power $p$ and reactive power $q$ setpoints for the inverters. To calculate the update, the \ac{OFO} controller solves a quadratic problem that is easy to solve even for large numbers of variables which means the setup scales very well with the system size. The updated $p$ and $q$ setpoints are then communicated to the inverters. Within a few iterations, $p$ and $q$ converge to the optimal solution of the overarching optimization problem.}

\section{Optimization Problems for Distribution Grid Flexibility}

The \ac{FPU} are supposed to deliver the requested flexibility $P_{set}$ at the \ac{PCC} while all voltages $v$, active power injections $p$, and reactive power injections $q$ are within their limits.
Generally, we can either encode this into the cost function or the constraints of the optimization problem~\eqref{eq:general_op_problem}. We refer to these two options as \emph{cost approach} and \emph{constraint approach}. Based on our experimental results, we can later on give a clear recommendation on which approach is favorable.

\subsection{Cost Approach}
In the cost approach, we encode the goal of providing flexibility in the cost function as follows

\begin{equation} \label{eq:cost_function_approach}
	\begin{aligned}
		\min_{p,q} \quad & (p_{set}-p_{PCC}(p,q))^2 \\
		\textrm{s. t.} \quad & p_{min} \leq p \leq p_{max} \\
        & q_{min} \leq q \leq q_{max} \\
        & v_{min} \leq v(p,q,d) \leq v_{max}
\end{aligned}
\end{equation}

The \ac{OFO} controller then is $u(k+1)=\begin{bmatrix}p(k+1)\\q(k+1)\end{bmatrix}=\begin{bmatrix}p(k)\\q(k)\end{bmatrix}+\sigma(u,d,y_m)$ with
\begin{equation} \label{eq:ofostepcalculation_cost}
	\begin{aligned}
		\sigma(u,d,y_m) \coloneqq \arg \min_{w \in \mathbb{R}^{n}} \, & \| w + 2G^{-1} H_p (p_{set}-p_{PCC}(p,q)))\|^{2}_G\\
		\textrm{s. t.} \quad & \begin{bmatrix}p_{min} \\q_{min}\end{bmatrix}\leq \begin{bmatrix}p(k) \\q(k)\end{bmatrix} + w \leq \begin{bmatrix}p_{max} \\q_{max}\end{bmatrix}\\
		\quad & v_{min} \leq v_{meas}(k) + H_v w \leq v_{max}
\end{aligned}
\end{equation}

\subsection{Constraint Approach}
In the constraint approach, we encode the goal of providing flexibility through the constraints $P_{set} = p_{PCC}$. Therefore, we can use the cost function to further specify what we consider the optimal behavior of our power grid. Here we consider a minimal use of active and reactive power to be optimal and therefore define
\begin{equation} \label{eq:constraint_approach}
	\begin{aligned}
		\min_{p,q} \quad & p^Tp+q^Tq \\
		\textrm{s. t.} \quad & p_{min} \leq p \leq p_{max} \\
        & q_{min} \leq q \leq q_{max} \\
        & v_{min} \leq v(p,q) \leq v_{max}\\
        & p_{set} = p_{PCC}(p,q)
\end{aligned}
\end{equation}

The \ac{OFO} controller then is $u(k+1) = \begin{bmatrix}p(k+1)\\q(k+1)\end{bmatrix}=\begin{bmatrix}p(k)\\q(k)\end{bmatrix}+\sigma(u,d,y_m)$ with
\begin{equation} \label{eq:ofostepcalculation_constraint}
	\begin{aligned}
		\sigma(u,d,y_m) \coloneqq \arg &\min_{w \in \mathbb{R}^{n}} \,  \| w + 2G^{-1}\begin{bmatrix}p(k)\\q(k)
		\end{bmatrix}\|^{2}_G\\
		\textrm{s. t.} \quad & \begin{bmatrix}p_{min} \\q_{min}\end{bmatrix}\leq \begin{bmatrix}p(k) \\q(k)\end{bmatrix} + w \leq \begin{bmatrix}p_{max} \\q_{max}\end{bmatrix}\\
		\quad & v_{min} \leq v_{meas} + H_v w \leq v_{max}\\
  		\quad & p_{set} = p_{PCC}(k)+ H_p w
\end{aligned}
\end{equation}

\section{Controller Tuning}

We will now have a closer look at how \ac{OFO} controllers work. This will become important when we aim to tune them through the matrix $G$. Note that, $G$ only affects the solution of the underlying \ac{QP} inside the \ac{OFO} controller but does not affect the solution of the overarching optimization problem\cite{haberle2020non}. In other words: The overarching optimization problem defines how the system is supposed to operate and $G$ can be used to adjust how the \ac{OFO} controller converges to this operating point. In the following we focus on tuning $G$.\\
Consider the \ac{QP} in~\eqref{eq:ofostepcalculation}. The cost function is a norm and therefore positive or zero. Without active constraints, the solution of the \ac{QP} can be chosen as
\begin{align}
    \begin{split}
        w&=-G^{-1}H(u)^{T}\nabla \Phi(u,y)\\
         &=-G^{-1}\nabla_u \Phi(h(u,d))
    \end{split}
\end{align}
and the cost function is then zero and therefore minimized.
This means, that if the constraints are not active, the controller integrates the gradient of the cost function of the overarching optimization problem multiplied with $G^{-1}$. Hence, this is a gradient descent scheme that keeps changing $u$ until the gradient is zero and has therefore converged to an unconstrained (local) minimum of~\eqref{eq:cost_function_approach}. The matrix $G^{-1}$ can be used to tune the convergence behavior. The smaller the entries in $G$, the faster the convergence.\\
The behavior is different when there are active constraints, e.g. a voltage reaches its limit. Then certain entries of $w$ need to deviate from the unconstrained solution to guarantee that the active constraint is not violated. Then the optimal solution of the \ac{QP} in~\eqref{eq:ofostepcalculation} is a vector $w$ that satisfies the constraints and is "as close as possible" to $-G^{-1}H(u)^{T}\nabla \Phi(u,y)$. Using the tuning matrix $G$ we can decide what close in this context means. This becomes more apparent when we multiply out the cost function of the \ac{QP}:
\begin{align*}
\begin{split}
    \| w + &G^{-1} \nabla_u \Phi(u)\|^{2}_G\\&= w^TGw + 2w^T\nabla_u \Phi(u) + \nabla_u \Phi(u)^T G^{-1} \nabla_u \Phi(u)
\end{split}
\end{align*}
We can see that the tuning matrix $G$ is a tuning weight in the term $w^TGw$. Therefore, small values of $G$ lead to a small weight on the corresponding entry in $w$, and the \ac{QP} can make these entries in $w$ large while still keeping the cost function of the \ac{QP} small. Therefore, if an element of $w$ has to deviate from the unconstrained optimal solution it will be these entries of $w$. Overall, $G$ gives us a way to tune which descent direction of the overarching optimization problem we care about most.

We will furthermore explain the tuning of $G$ by having a look at the cost approach and the resulting \ac{OFO} controller in~\eqref{eq:ofostepcalculation_cost}.
Figure~\ref{fig:slow_cost} shows the convergence behavior of the \ac{OFO} controller with $G=I$ for the cost approach.
\begin{figure}
    \centering
    \setlength\fwidth{\columnwidth}
	\setlength\fheight{15cm}
    \begin{adjustbox}{max width=\columnwidth}
    \input{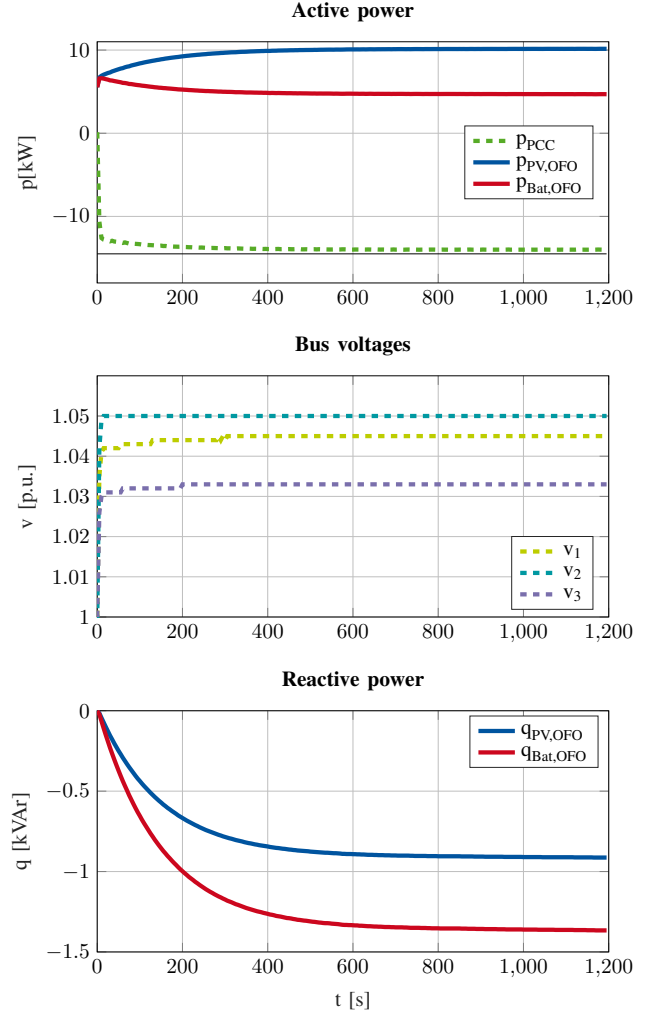}
\end{adjustbox}
    \caption{Extremely slow convergence of the cost approach with $G=I$.}
    \label{fig:slow_cost}
\end{figure}
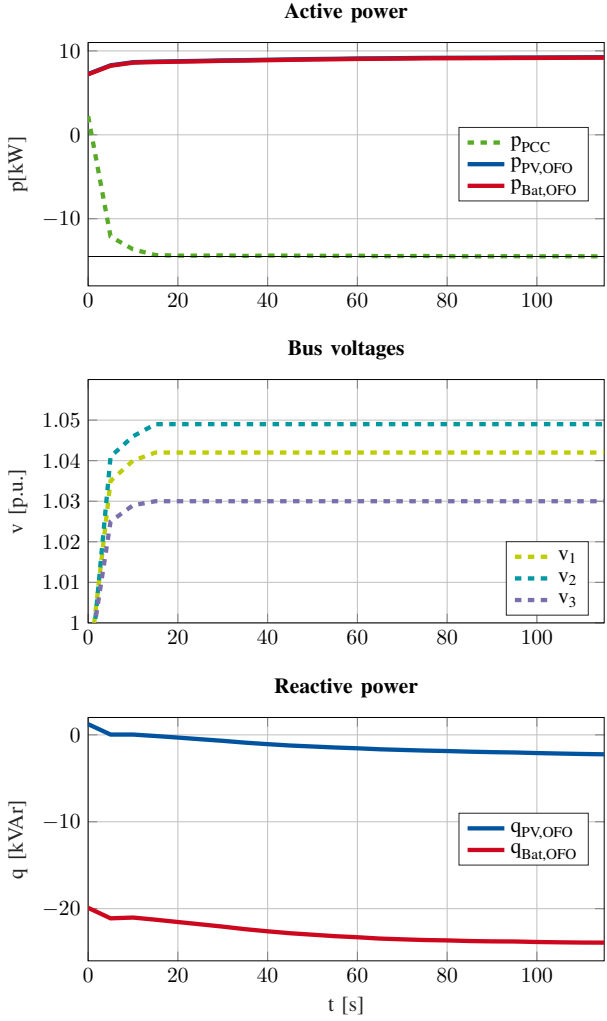
\begin{figure}
    \centering
    \setlength\fwidth{\columnwidth}
	\setlength\fheight{15cm}
    \begin{adjustbox}{max width=\columnwidth}
%
%
\definecolor{mycolor1}{rgb}{0.34118,0.67059,0.15294}%
\definecolor{mycolor2}{rgb}{0.00000,0.32941,0.62353}%
\definecolor{mycolor3}{rgb}{0.80000,0.02745,0.11765}%
\definecolor{mycolor4}{rgb}{0.74118,0.80392,0.00000}%
\definecolor{mycolor5}{rgb}{0.00000,0.59608,0.63137}%
\definecolor{mycolor6}{rgb}{0.47843,0.43529,0.67451}%
\begin{tikzpicture}

\begin{axis}[%
width=0.951\fwidth,
height=0.265\fheight,
at={(0\fwidth,0.735\fheight)},
scale only axis,
xmin=0.000,
xmax=115.000,
ymin=-18.000,
ymax=11.000,
ylabel style={font=\color{white!15!black}},
ylabel={p[kW]},
axis background/.style={fill=white},
title style={font=\bfseries},
title={Active power},
xmajorgrids,
ymajorgrids,
legend style={at={(0.97,0.5)}, anchor=east, legend cell align=left, align=left, draw=white!15!black}
]
\addplot [color=mycolor1, dashed, line width=2.0pt]
  table[row sep=crcr]{%
0.000	2.230\\
5.000	-12.123\\
10.000	-13.643\\
15.000	-14.330\\
20.000	-14.399\\
25.000	-14.402\\
30.000	-14.366\\
35.000	-14.431\\
40.000	-14.394\\
45.000	-14.399\\
50.000	-14.412\\
55.000	-14.430\\
60.000	-14.404\\
65.000	-14.462\\
70.000	-14.445\\
75.000	-14.437\\
80.000	-14.475\\
85.000	-14.480\\
90.000	-14.479\\
95.000	-14.467\\
100.000	-14.486\\
105.000	-14.469\\
110.000	-14.476\\
115.000	-14.483\\
};
\addlegendentry{$\text{p}_{\text{PCC}}$}

\addplot [color=mycolor2, line width=2.0pt]
  table[row sep=crcr]{%
0.000	7.251\\
5.000	8.281\\
10.000	8.653\\
15.000	8.726\\
20.000	8.770\\
25.000	8.813\\
30.000	8.871\\
35.000	8.901\\
40.000	8.947\\
45.000	8.990\\
50.000	9.029\\
55.000	9.059\\
60.000	9.101\\
65.000	9.117\\
70.000	9.141\\
75.000	9.168\\
80.000	9.179\\
85.000	9.188\\
90.000	9.197\\
95.000	9.212\\
100.000	9.218\\
105.000	9.231\\
110.000	9.242\\
115.000	9.250\\
};
\addlegendentry{$\text{p}_{\text{PV,OFO}}$}

\addplot [color=mycolor3, line width=2.0pt]
  table[row sep=crcr]{%
0.000	7.207\\
5.000	8.231\\
10.000	8.600\\
15.000	8.673\\
20.000	8.717\\
25.000	8.759\\
30.000	8.817\\
35.000	8.846\\
40.000	8.892\\
45.000	8.936\\
50.000	8.974\\
55.000	9.004\\
60.000	9.045\\
65.000	9.062\\
70.000	9.085\\
75.000	9.113\\
80.000	9.123\\
85.000	9.132\\
90.000	9.141\\
95.000	9.156\\
100.000	9.162\\
105.000	9.175\\
110.000	9.186\\
115.000	9.193\\
};
\addlegendentry{$\text{p}_{\text{Bat,OFO}}$}

\addplot [color=black, forget plot]
  table[row sep=crcr]{%
0.000	-14.500\\
115.000	-14.500\\
};
\end{axis}

\begin{axis}[%
width=0.951\fwidth,
height=0.265\fheight,
at={(0\fwidth,0.368\fheight)},
scale only axis,
xmin=0.000,
xmax=115.000,
ymin=1.000,
ymax=1.060,
ytick={   1, 1.01, 1.02, 1.03, 1.04, 1.05},
ylabel style={font=\color{white!15!black}},
ylabel={v [p.u.]},
axis background/.style={fill=white},
title style={font=\bfseries},
title={Bus voltages},
xmajorgrids,
ymajorgrids,
legend style={at={(0.97,0.03)}, anchor=south east, legend cell align=left, align=left, draw=white!15!black}
]
\addplot [color=mycolor4, dashed, line width=2.0pt]
  table[row sep=crcr]{%
0.000	0.987\\
5.000	1.035\\
10.000	1.040\\
15.000	1.042\\
20.000	1.042\\
25.000	1.042\\
30.000	1.042\\
35.000	1.042\\
40.000	1.042\\
45.000	1.042\\
50.000	1.042\\
55.000	1.042\\
60.000	1.042\\
65.000	1.042\\
70.000	1.042\\
75.000	1.042\\
80.000	1.042\\
85.000	1.042\\
90.000	1.042\\
95.000	1.042\\
100.000	1.042\\
105.000	1.042\\
110.000	1.042\\
115.000	1.042\\
};
\addlegendentry{$\text{v}_{\text{1}}$}

\addplot [color=mycolor5, dashed, line width=2.0pt]
  table[row sep=crcr]{%
0.000	0.982\\
5.000	1.041\\
10.000	1.046\\
15.000	1.049\\
20.000	1.049\\
25.000	1.049\\
30.000	1.049\\
35.000	1.049\\
40.000	1.049\\
45.000	1.049\\
50.000	1.049\\
55.000	1.049\\
60.000	1.049\\
65.000	1.049\\
70.000	1.049\\
75.000	1.049\\
80.000	1.049\\
85.000	1.049\\
90.000	1.049\\
95.000	1.049\\
100.000	1.049\\
105.000	1.049\\
110.000	1.049\\
115.000	1.049\\
};
\addlegendentry{$\text{v}_{\text{2}}$}

\addplot [color=mycolor6, dashed, line width=2.0pt]
  table[row sep=crcr]{%
0.000	0.989\\
5.000	1.025\\
10.000	1.029\\
15.000	1.030\\
20.000	1.030\\
25.000	1.030\\
30.000	1.030\\
35.000	1.030\\
40.000	1.030\\
45.000	1.030\\
50.000	1.030\\
55.000	1.030\\
60.000	1.030\\
65.000	1.030\\
70.000	1.030\\
75.000	1.030\\
80.000	1.030\\
85.000	1.030\\
90.000	1.030\\
95.000	1.030\\
100.000	1.030\\
105.000	1.030\\
110.000	1.030\\
115.000	1.030\\
};
\addlegendentry{$\text{v}_{\text{3}}$}

\end{axis}

\begin{axis}[%
width=0.951\fwidth,
height=0.265\fheight,
at={(0\fwidth,0\fheight)},
scale only axis,
xmin=0.000,
xmax=115.000,
xlabel style={font=\color{white!15!black}},
xlabel={t [s]},
ymin=-26.000,
ymax=2.000,
ylabel style={font=\color{white!15!black}},
ylabel={q [kVAr]},
axis background/.style={fill=white},
title style={font=\bfseries},
title={Reactive power},
xmajorgrids,
ymajorgrids,
legend style={at={(0.97,0.5)}, anchor=east, legend cell align=left, align=left, draw=white!15!black}
]
\addplot [color=mycolor2, line width=2.0pt]
  table[row sep=crcr]{%
0.000	1.233\\
5.000	0.035\\
10.000	0.034\\
15.000	-0.132\\
20.000	-0.312\\
25.000	-0.502\\
30.000	-0.686\\
35.000	-0.896\\
40.000	-1.069\\
45.000	-1.225\\
50.000	-1.338\\
55.000	-1.454\\
60.000	-1.544\\
65.000	-1.659\\
70.000	-1.734\\
75.000	-1.808\\
80.000	-1.861\\
85.000	-1.934\\
90.000	-1.988\\
95.000	-2.027\\
100.000	-2.094\\
105.000	-2.148\\
110.000	-2.198\\
115.000	-2.236\\
};
\addlegendentry{$\text{q}_{\text{PV,OFO}}$}

\addplot [color=mycolor3, line width=2.0pt]
  table[row sep=crcr]{%
0.000	-19.902\\
5.000	-21.108\\
10.000	-21.021\\
15.000	-21.265\\
20.000	-21.533\\
25.000	-21.796\\
30.000	-22.063\\
35.000	-22.358\\
40.000	-22.603\\
45.000	-22.823\\
50.000	-22.993\\
55.000	-23.162\\
60.000	-23.285\\
65.000	-23.439\\
70.000	-23.525\\
75.000	-23.606\\
80.000	-23.649\\
85.000	-23.722\\
90.000	-23.760\\
95.000	-23.774\\
100.000	-23.833\\
105.000	-23.868\\
110.000	-23.896\\
115.000	-23.905\\
};
\addlegendentry{$\text{q}_{\text{Bat,OFO}}$}

\end{axis}

\begin{axis}[%
width=1.227\fwidth,
height=1.227\fheight,
at={(-0.16\fwidth,-0.135\fheight)},
scale only axis,
xmin=0.000,
xmax=1.000,
ymin=0.000,
ymax=1.000,
axis line style={draw=none},
ticks=none,
axis x line*=bottom,
axis y line*=left
]
\end{axis}
\end{tikzpicture}%
\end{adjustbox}
    \caption{Fast convergence of the cost approach due to a tuned $G$ matrix.}
    \label{fig:fast_cost}
\end{figure}
In the beginning, the controller quickly changes the active power setpoints to track $P_{set}=-14.5$~kW while the reactive power setpoints stay small. Then the convergence becomes very slow once the voltage at the BESS inverter (bus~2) reaches its upper level (dashed turquoise line). From this point onward a voltage constraint in the \ac{QP} in~\eqref{eq:ofostepcalculation_cost} is active. Changing the active power injections to further minimize the cost function of the overarching optimization problem~\eqref{eq:cost_function_approach} is now only possible if reactive power is used to counteract the voltage rise due to the active power increase. Because a change in the reactive powers $q$ will not affect the cost function ($\nabla_q p_{PCC}\approx0$), the unconstrained solution for $w$ is to not change the reactive power injections. Therefore, the \ac{QP} does not want to change $q$. However, without changing $q$ we cannot change $p$ because that would lead to voltage violations. By putting small values in the entries of $G$ corresponding to $q$ the \ac{QP} can change $q$ without a large effect on the cost function of the \ac{QP}. Then the optimal solution of the \ac{QP} will be to make larger changes to the reactive power injections which allow for larger changes in the active power injections $p$. Figure~\ref{fig:fast_cost} shows the behavior of the cost approach with \begin{equation*}
    G=\begin{bmatrix}I_2 &0\\0 &10^{-10}\cdot I_2\end{bmatrix},
\end{equation*}
where $I_2$ is an identity matrix of size $2\times2$.
The figure shows that the convergence is significantly improved because the reactive power injections $q$ are changed faster which allows for faster changes in $p$ and therefore convergence of $p_{PCC}$ to the flexibility setpoint $p_{set}$. Within three iterations of the \ac{OFO} controller, the distribution grid successfully delivers the requested flexibility to the upper-level grid. Furthermore, the constraints within the distribution grid are satisfied. We conclude that $G$ must be used to tune the convergence behavior of an \ac{OFO} controller. When constraints become active then tuning $G$ is even more important, because otherwise the convergence speed can be significantly affected.

\section{Results}
In this section, we present the results of the \emph{constraint approach} and compare it to the behavior of the \emph{cost approach}. Furthermore, we show the interaction of an \ac{OFO} controller with an inverter in the grid that is not part of the \ac{OFO} control but performs voltage control according to VDE4105~\cite{VDE}. Finally, we also add a charging electric vehicle which enables us to test our \ac{OFO} controller on a real grid with real loads.
\begin{figure}
    \centering
    \setlength\fwidth{\columnwidth}
	\setlength\fheight{15cm}
    \begin{adjustbox}{max width=\columnwidth}
%
%
\definecolor{mycolor1}{rgb}{0.34118,0.67059,0.15294}%
\definecolor{mycolor2}{rgb}{0.00000,0.32941,0.62353}%
\definecolor{mycolor3}{rgb}{0.80000,0.02745,0.11765}%
\definecolor{mycolor4}{rgb}{0.74118,0.80392,0.00000}%
\definecolor{mycolor5}{rgb}{0.00000,0.59608,0.63137}%
\definecolor{mycolor6}{rgb}{0.47843,0.43529,0.67451}%
\begin{tikzpicture}

\begin{axis}[%
width=0.951\fwidth,
height=0.265\fheight,
at={(0\fwidth,0.735\fheight)},
scale only axis,
xmin=0.000,
xmax=25.000,
ymin=-18.000,
ymax=14.000,
ylabel style={font=\color{white!15!black}},
ylabel={p[kW]},
axis background/.style={fill=white},
title style={font=\bfseries},
title={Active power},
xmajorgrids,
ymajorgrids,
legend style={at={(0.97,0.5)}, anchor=east, legend cell align=left, align=left, draw=white!15!black}
]
\addplot [color=mycolor1, dashed, line width=2.0pt]
  table[row sep=crcr]{%
0.000	0.123\\
5.000	-14.023\\
10.000	-14.434\\
15.000	-14.506\\
20.000	-14.502\\
25.000	-14.504\\
30.000	-14.482\\
35.000	-14.510\\
40.000	-14.512\\
45.000	-14.495\\
50.000	-14.501\\
55.000	-14.493\\
60.000	-14.494\\
65.000	-14.509\\
70.000	-14.497\\
75.000	-14.548\\
80.000	-14.451\\
85.000	-14.508\\
90.000	-14.494\\
95.000	-14.503\\
100.000	-14.496\\
105.000	-14.505\\
110.000	-14.501\\
115.000	-14.495\\
};
\addlegendentry{$\text{p}_{\text{PCC}}$}

\addplot [color=mycolor2, line width=2.0pt]
  table[row sep=crcr]{%
0.000	-0.000\\
5.000	10.984\\
10.000	11.884\\
15.000	11.415\\
20.000	11.462\\
25.000	11.456\\
30.000	11.434\\
35.000	11.409\\
40.000	11.380\\
45.000	11.410\\
50.000	11.398\\
55.000	11.441\\
60.000	11.435\\
65.000	11.441\\
70.000	11.408\\
75.000	11.463\\
80.000	11.435\\
85.000	11.446\\
90.000	11.432\\
95.000	11.456\\
100.000	11.389\\
105.000	11.443\\
110.000	11.406\\
115.000	11.485\\
};
\addlegendentry{$\text{p}_{\text{PV,OFO}}$}

\addplot [color=mycolor3, line width=2.0pt]
  table[row sep=crcr]{%
0.000	-0.000\\
5.000	3.886\\
10.000	3.467\\
15.000	4.007\\
20.000	3.954\\
25.000	3.957\\
30.000	3.975\\
35.000	4.018\\
40.000	4.037\\
45.000	3.994\\
50.000	4.011\\
55.000	3.966\\
60.000	3.980\\
65.000	3.980\\
70.000	4.004\\
75.000	3.952\\
80.000	3.931\\
85.000	3.970\\
90.000	3.976\\
95.000	3.958\\
100.000	4.022\\
105.000	3.973\\
110.000	4.004\\
115.000	3.924\\
};
\addlegendentry{$\text{p}_{\text{Bat,OFO}}$}

\addplot [color=black, forget plot]
  table[row sep=crcr]{%
0.000	-14.500\\
115.000	-14.500\\
};
\end{axis}

\begin{axis}[%
width=0.951\fwidth,
height=0.265\fheight,
at={(0\fwidth,0.368\fheight)},
scale only axis,
xmin=0.000,
xmax=25.000,
ymin=1.000,
ymax=1.060,
ytick={   1, 1.01, 1.02, 1.03, 1.04, 1.05},
ylabel style={font=\color{white!15!black}},
ylabel={v [p.u.]},
axis background/.style={fill=white},
title style={font=\bfseries},
title={Bus voltages},
xmajorgrids,
ymajorgrids,
legend style={at={(0.97,0.03)}, anchor=south east, legend cell align=left, align=left, draw=white!15!black}
]
\addplot [color=mycolor4, dashed, line width=2.0pt]
  table[row sep=crcr]{%
0.000	1.000\\
5.000	1.045\\
10.000	1.046\\
15.000	1.046\\
20.000	1.046\\
25.000	1.046\\
30.000	1.046\\
35.000	1.046\\
40.000	1.046\\
45.000	1.046\\
50.000	1.046\\
55.000	1.046\\
60.000	1.046\\
65.000	1.046\\
70.000	1.046\\
75.000	1.046\\
80.000	1.046\\
85.000	1.046\\
90.000	1.046\\
95.000	1.046\\
100.000	1.046\\
105.000	1.046\\
110.000	1.046\\
115.000	1.046\\
};
\addlegendentry{$\text{v}_{\text{1}}$}

\addplot [color=mycolor5, dashed, line width=2.0pt]
  table[row sep=crcr]{%
0.000	0.999\\
5.000	1.049\\
10.000	1.049\\
15.000	1.050\\
20.000	1.050\\
25.000	1.050\\
30.000	1.050\\
35.000	1.050\\
40.000	1.050\\
45.000	1.050\\
50.000	1.050\\
55.000	1.050\\
60.000	1.050\\
65.000	1.050\\
70.000	1.050\\
75.000	1.050\\
80.000	1.050\\
85.000	1.050\\
90.000	1.050\\
95.000	1.050\\
100.000	1.050\\
105.000	1.050\\
110.000	1.050\\
115.000	1.050\\
};
\addlegendentry{$\text{v}_{\text{2}}$}

\addplot [color=mycolor6, dashed, line width=2.0pt]
  table[row sep=crcr]{%
0.000	1.000\\
5.000	1.033\\
10.000	1.034\\
15.000	1.034\\
20.000	1.034\\
25.000	1.034\\
30.000	1.034\\
35.000	1.034\\
40.000	1.034\\
45.000	1.034\\
50.000	1.034\\
55.000	1.034\\
60.000	1.034\\
65.000	1.034\\
70.000	1.034\\
75.000	1.034\\
80.000	1.034\\
85.000	1.034\\
90.000	1.034\\
95.000	1.034\\
100.000	1.034\\
105.000	1.034\\
110.000	1.034\\
115.000	1.034\\
};
\addlegendentry{$\text{v}_{\text{3}}$}

\end{axis}

\begin{axis}[%
width=0.951\fwidth,
height=0.265\fheight,
at={(0\fwidth,0\fheight)},
scale only axis,
xmin=0.000,
xmax=25.000,
xlabel style={font=\color{white!15!black}},
xlabel={t [s]},
ymin=-2.500,
ymax=0.000,
ylabel style={font=\color{white!15!black}},
ylabel={q [kVAr]},
axis background/.style={fill=white},
title style={font=\bfseries},
title={Reactive power},
xmajorgrids,
ymajorgrids,
legend style={legend cell align=left, align=left, draw=white!15!black}
]
\addplot [color=mycolor2, line width=2.0pt]
  table[row sep=crcr]{%
0.000	-0.000\\
5.000	-1.190\\
10.000	-1.412\\
15.000	-1.242\\
20.000	-1.259\\
25.000	-1.258\\
30.000	-1.251\\
35.000	-1.239\\
40.000	-1.231\\
45.000	-1.243\\
50.000	-1.239\\
55.000	-1.253\\
60.000	-1.250\\
65.000	-1.251\\
70.000	-1.241\\
75.000	-1.260\\
80.000	-1.258\\
85.000	-1.254\\
90.000	-1.250\\
95.000	-1.257\\
100.000	-1.235\\
105.000	-1.253\\
110.000	-1.241\\
115.000	-1.268\\
};
\addlegendentry{$\text{q}_{\text{PV,OFO}}$}

\addplot [color=mycolor3, line width=2.0pt]
  table[row sep=crcr]{%
0.000	-0.000\\
5.000	-1.780\\
10.000	-2.113\\
15.000	-1.858\\
20.000	-1.884\\
25.000	-1.881\\
30.000	-1.871\\
35.000	-1.854\\
40.000	-1.842\\
45.000	-1.860\\
50.000	-1.853\\
55.000	-1.875\\
60.000	-1.870\\
65.000	-1.872\\
70.000	-1.857\\
75.000	-1.884\\
80.000	-1.882\\
85.000	-1.875\\
90.000	-1.871\\
95.000	-1.881\\
100.000	-1.848\\
105.000	-1.874\\
110.000	-1.857\\
115.000	-1.897\\
};
\addlegendentry{$\text{q}_{\text{Bat,OFO}}$}

\end{axis}
\end{tikzpicture}%
\end{adjustbox}
    \caption{Control performance of the constraint approach.}
    \label{fig:conststrained}
\end{figure}
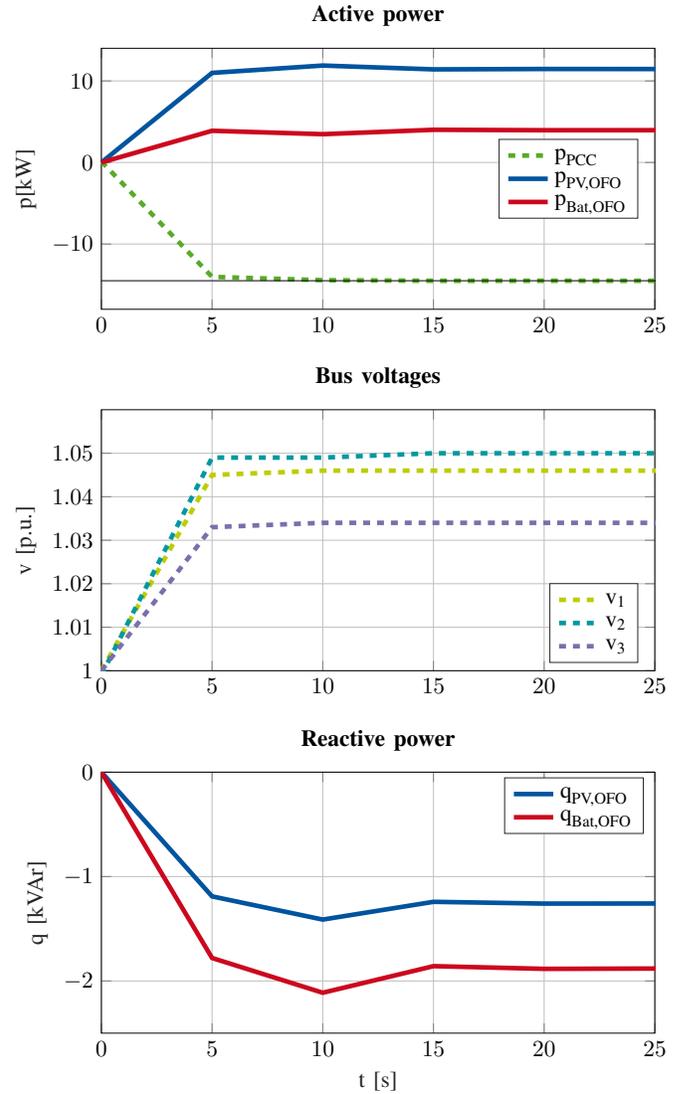
\subsection{Performance of the Constraint Approach}
In Figure~\ref{fig:conststrained} it can be seen that the convergence of the constraint approach is nearly immediate. After 10 seconds, which is two iterations of the controller, the setpoint is tracked. Again, all constraints are satisfied and therefore the grid is providing the requested flexibility while operating the distribution grid within its operational limits. The convergence is even faster than with the cost approach but the main difference is the use of reactive power. In the cost approach, there is no penalty on the use of reactive power leading to a high reactive power usage. In the constraint approach, the usage of reactive power is part of the cost function of the overarching control problem~\eqref{eq:constraint_approach} and therefore the \ac{OFO} controller is encouraged to limit the use of reactive power. This leads to the following behavior that can be seen in Figure~\ref{fig:conststrained}: The active power setpoints for the two inverters are not the same. The inverter at the end of the line is utilized less because its active power injection leads to a voltage increase at its point of connection (bus~2) that would need to be compensated by using reactive power. Given that these reactive power flows lead to active power losses it is advantageous to limit the use of reactive power and therefore losses. With the constraint approach, the losses are significantly lower than with the cost approach. This becomes apparent when we compare the active power injections of the inverters. To deliver -14.5~kW of flexibility to the upper-level grid the inverters change their active power injections by $-4-11=-15$~kW in the constraint approach and by $-9.2-9.2=-18.4$~kW in the cost approach. This highly non-trivial solution of utilizing inverters differently depending on where they are located in the grid is a remarkable behavior of the controller which is able to derive this solution solely based on the approximate sensitivities $H_p$ and $H_v$ and feedback from the grid through measurements. Also note that, if the distribution grid is not able to provide the requested flexibility the \ac{QP} in the constraint approach becomes infeasible. This is because $p_{set}=p_{PCC}$ is a hard constraint of the QP. This infeasibility can be used as a signal to communicated to the entity requesting the flexibility that this is currently not possible. The \ac{QP} in the cost approach does not become infeasible because the goal of $p_{set}=p_{PCC}$ is included in the cost function and does therefore not have this feature.\\
Overall, the constraint approach is faster and uses the inverters more efficiently.

\subsection{Interaction with other Inverters in the Grid}
\begin{figure}
    \centering
    \setlength\fwidth{\columnwidth}
	\setlength\fheight{15cm}
    \begin{adjustbox}{max width=1\columnwidth}
    \input{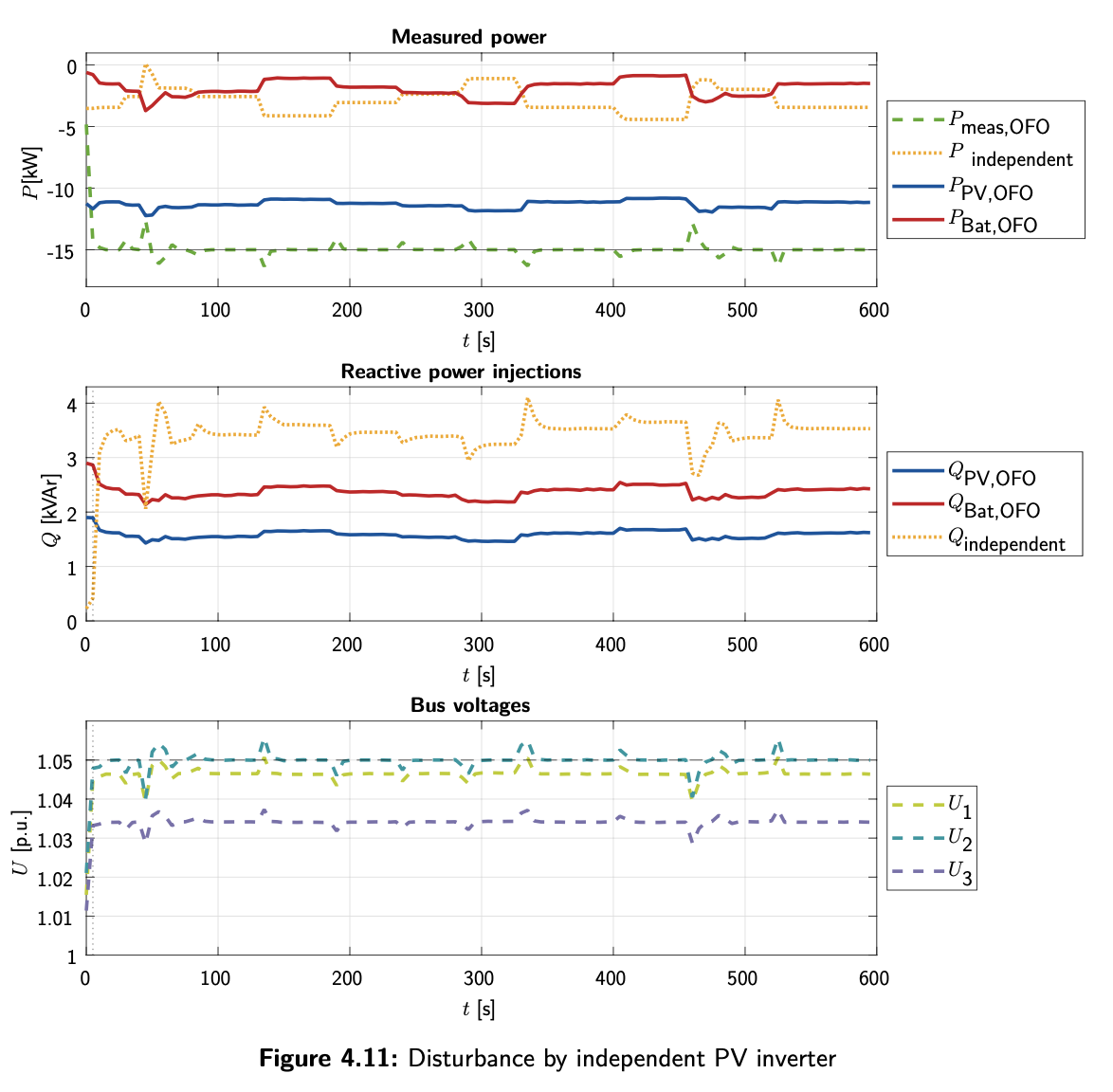}
\end{adjustbox}
    \caption{Interaction of the \ac{OFO} controller with an inverter injecting a varying amount of power and which is controlling its reactive power according to VDE4105~\cite{VDE}.}
    \label{fig:legacy_inverter}
\end{figure}

We will now show and analyze how the \ac{OFO} controller with the constraint approach interacts with other devices in the grid. For that, we add the PV-Inverter~II to the setup. This inverter is injecting a changing amount of active power into the grid and operates a $q(v)$ voltage control according to VDE4105~\cite{VDE} with a deadband of 3\%.\\
The task is to provide a flexibility of -15~kW at the \ac{PCC}. The dashed orange line in Figure~\ref{fig:legacy_inverter} shows that deviations from the setpoint are immediately mitigated in the next time step, meaning once the controller detects a deviation it fixes it within one sampling interval and accurately. Figure~\ref{fig:legacy_inverter} shows that small voltage violations can occur when the independent PV inverter~II changes its active power injections. Also, these violations are mitigated within one time step of the controller which is well within the timespan of the current norms~\cite{VDE}. Overall, the \ac{OFO} controller interacts well with the other inverter in the grid.

\subsection{Fully Realistic Test Setup}
\begin{figure}
    \centering
    \setlength\fwidth{\columnwidth}
	\setlength\fheight{15cm}
    \begin{adjustbox}{max width=\columnwidth}
    \input{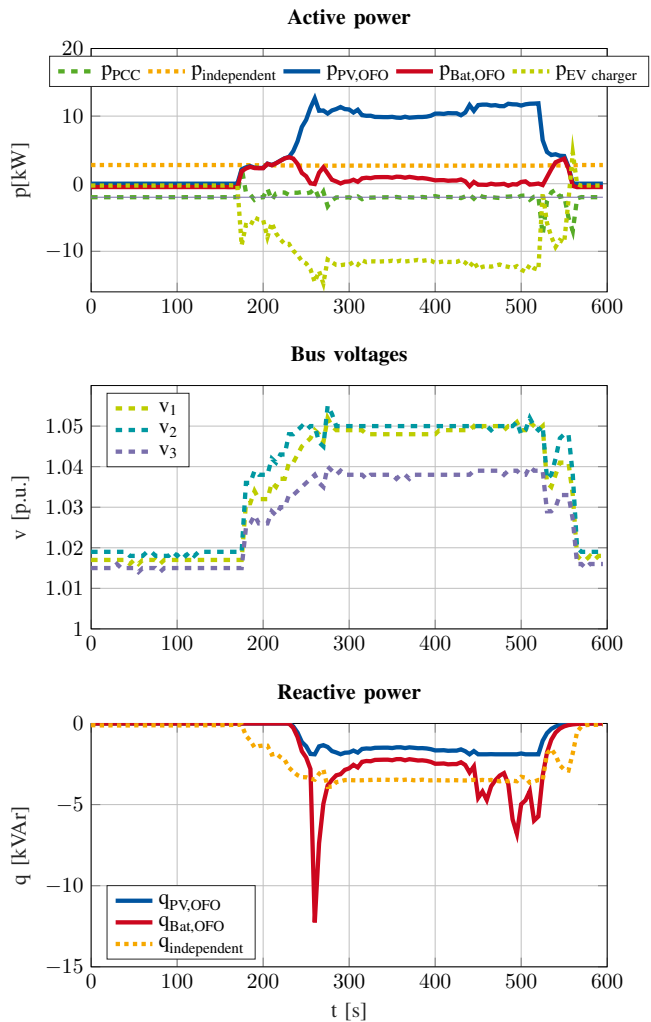}
\end{adjustbox}
    \caption{An electric vehicle is charged and the \ac{OFO} controller counteracts this disturbance such that the distribution grid accurately provides the requested flexibility while satisfying the distribution grid constraints.}
    \label{fig:EV_charging}
\end{figure}

To go one step further, we add the EV charging point to the grid and connect an electric vehicle. The requested flexibility setpoint at the \ac{PCC} is now -2~kW. The collected data can be seen in Figure~\ref{fig:EV_charging}. The setpoint is tracked well. The only noticeable deviations are in the beginning and end when the EV continuously changes its active power intake. This is because the \ac{OFO} controller is not aware of the changing power consumption of the EV. Actually, the \ac{OFO} controller does not even know that there is an EV charging point in the grid. Therefore, the \ac{OFO} controller can only react to the effect of the changing power consumption which leads to a small tracking error. However, this is not specific to the \ac{OFO} control method but the ramp-like disturbance that the charging EV represents. The voltages are kept within their limits and violations are mitigated quickly.

\section{Conclusion}
In this paper, we used an \ac{OFO} controller to disaggregate a flexibility request onto \ac{FPU}s. Using a real distribution grid setup we showed that tuning the matrix $G$ of the \ac{OFO} controller is essential for fast convergence and good performance and we explained how to tune $G$ in detail. We compared two possible approaches to phrase the disaggregation task as an optimization problem and can conclude that the constraint approach outperforms the cost approach. The better performance is in terms of losses in the grid and the constrained approach is more versatile as its cost function can be used to parameterize the favored behavior of the system. Also the constraint approach will realize if the distribution grid cannot provide the requested flexibility and can signal this to the entity requesting the flexibility. Furthermore, the experiments showed that the tuned \ac{OFO} controller is compatible with other control devices in the grid and enables fast delivery of the requested flexibility while guaranteeing the operational limits in the distribution grid. Also, the method is robust to model mismatch and changing consumption in the distribution grid while it is able to utilize the full flexibility of the distribution grid. Overall, \ac{OFO} with the constraint approach is a powerful tool to disaggregate flexibility requests onto \ac{FPU}s in a distribution grid.

\bibliographystyle{IEEEtran}
\bibliography{IEEEabrv,biblio}

\end{document}